\documentstyle[aps]{revtex}
\begin{document}
\title{Phases of a Type Ia supernova explosion}

\author{Jens C. Niemeyer\footnote{E-mail: j-niemeyer@uchicago.edu,
WWW: http://bigwhirl.uchicago.edu/jcn/}}

\address{University of Chicago, Department of Astronomy and
Astrophysics, 5640 S. Ellis Avenue, Chicago, IL 60637}

\maketitle

\begin{abstract}

In the framework of the Chandrasekhar mass white dwarf model for Type
Ia supernovae, various stages of the explosion are described in
terms of the burning regimes of the thermonuclear flame
front. In the early flamelet regime following the ``smoldering'' phase
prior to the explosion, the flame is sufficiently thin and fast to
remain laminar on small scales. As the white dwarf density declines,
the thermal flame structure becomes subject to penetration by
turbulent eddies, and it enters the ``distributed burning'' regime. A
specific control parameter for this transition is
proposed. Furthermore, we outline an argument for the coincidence of
the transition between burning regimes with the onset of a
deflagration-detonation-transition (DDT) in the late phase of the
explosion.
\end{abstract}

\section{Introduction}
Theoretical efforts to understand the physics of Type Ia supernovae
(SN Ia's) based on the model of exploding Chandrasekhar mass white
dwarfs have revealed a fascinating degree of
complexity. While the basic model appears simple -- a carbon and
oxygen white dwarf that, by some means, reaches critical conditions
for thermonuclear runaway inevitably burns a sizeable fraction of its mass 
to $^{56}$Ni, releasing roughly the amount of energy needed to power
the explosion ($\sim 10^{51}$ ergs) -- the hydrodynamics of the
explosion is far from being fully understood. As is well known, the
observed abundance of 
intermediate mass elements in the ejecta rules out the possibility that
the star is disrupted by a single, prompt detonation (a shock-driven
combustion wave), which would burn almost the entire star 
to nuclear statistical equilibrium \cite{arn69}. Consequently, any
successful model must involve an initial -- conductively propagating and
thus subsonic -- deflagration, or ``flame'', 
phase that allows the star to pre-expand. Turbulence driven by
buoyancy of the burnt material with respect to the unburnt background
increases the total burning rate, first by wrinkling the flame
surface and later by directly mixing hot ashes with cold fuel. Only allowed to
occur below a critical density of $\sim 10^{7}$ g cm$^{-3}$, the process
of direct turbulent mixing might be responsible for triggering a
detonation in the late stage of the explosion or during
the re-collapse of the partly burned white dwarf \cite{kho97,niewoo97}. 

The complicated interaction of 
thermonuclear burning, thermal conduction, turbulent mixing, and
buoyancy forces on scales covering up to 12 orders of magnitude makes
a full numerical solution of the governing equations virtually
impossible. Multidimensional simulations on the scale of the white
dwarf radius suffer from insufficient resolution of small scale
turbulence and of the thermonuclear flame front, and thus (implicitly or
explicitly) involve subgrid modeling to some extent, rendering the
interpretation of their results intrinsically difficult. One of the
central assumptions made in previous calculations is the notion of a
thin flame surface separating fuel (in our case, carbon and oxygen)
and ashes (nickel), and propagating into the
fuel at a speed that only depends on its density and
composition. Under certain conditions, the
laminar flame structure remains unperturbed by turbulent eddies on
small scales, while at the same time turbulence wrinkles or even
fragments the flame surface on larger scales. This so-called
``corrugated flamelet regime'' \cite{pet88} occurs early during the
explosion, when the 
flame is fast and thin and turbulence is weak. Key parameters for
large-scale simulations in the flamelet regime are the boundaries of
validity of the flamelet assumption, the small scale cut-off for flame
surface perturbations, and the turbulent flame speed
(the global propagation rate of the turbulent flame brush) as a
function of the laminar flame speed and 
turbulence intensity. Section (\ref{flamelet}) gives an overview
of thermonuclear burning in the flamelet regime.

It has recently become obvious that one cannot
ignore the effects of turbulence on the laminar flame structure once
the star has expanded to densities below $\sim 10^{7}$ g cm$^{-3}$
\cite{nieker1}. As soon as turbulent eddies on the scale of the flame
thickness become sufficiently strong to disrupt the thermal flame
structure and
mix fresh fuel into the hot burning products, the flame enters the
``distributed burning regime'' \cite{pope87}. It has been noted \cite{niewoo97}
that this process generates favorable conditions for a
deflagration-detonation-transition (DDT) that is thought to occur in
many SN Ia events \cite{hoe96}. Clearly, an explicit control parameter that
describes the transition from flamelet to distributed burning would be
useful; one possible choice, the ratio of turbulent and conductive
diffusivities, will be discussed in section (\ref{dist}). 

The following three sections summarize the general
features of the various stages of a Chandrasekhar mass SN Ia explosion,
described in terms of their respective burning regimes. It should be
kept in mind that the relations sketched below are chiefly based 
on dimensional and scaling arguments. Wherever feasible, accurate
numerical simulations will have to be carried out in order to verify
or discard these arguments and to specify numerical coefficients.

\section{Phase 1: Smoldering and ignition}
Very little is known about the time between the onset of the thermonuclear
runaway and the formation of the flame itself. At a central
temperature of $T \approx 2 \times 10^8$ K, neutrino cooling fails to keep up
with nuclear heating, and the core region begins to
``smolder''\cite{woowea94}. During the following $\sim$ 1000 years, the
core experiences internally heated convection with progressively
smaller turnover time scales $\tau_{\rm c} \sim \nu_{\rm bv}^{-1}$,
where $\nu_{\rm bv}$ is the Brunt-Vaisala frequency. Simultaneously, the
typical time scale for thermonuclear burning, $\tau_{\rm b}$, drops
even faster as a result of the rising core temperature and the steep
temperature dependence of the nuclear reaction rates. At $T \approx 6
\times 10^8$~K, both time scales become comparable, indicating that
convective plumes burn at
the same rate as they circulate \cite{woo90}. Experimental or numerical data
describing this regime of strong reactive convection is not available.

At {$T \approx 1.5 \times
10^9$ K}, $\tau_{\rm b}$ becomes extremely small compared with
$\tau_{\rm c}$, and carbon and oxygen
virtually burn in place. A new equilibrium between energy generation
and transport is found on much smaller length scales, $l \approx
10^{-4}$ cm, where thermal conduction by degenerate electrons balances
nuclear energy input \cite{timwoo92}. The flame is born.

The evolution of the runaway immediately prior to ignition of
the flame is crucial for determining its initial location and
shape. Using a simple toy model, Garcia \& Woosley \cite{garwoo95}
found that under certain conditions, burning bubbles subject to
buoyancy and drag forces can rise appreciable distances before flame
formation, suggesting the possibility of off-center ignition. As a
consequence, more material burns  at lower densities, thus
producing higher amounts of intermediate mass elements than a centrally
ignited explosion. A parameter study demonstrated the significant
influence of the location and number of initially ignited spots on
the final explosion energetics and nucleosynthesis \cite{nieea96}. 

The topology of the initial
flame surface is directly linked to the spatial and spectral state
of temperature fluctuations due to the strongly coupled dynamics of
three-dimensional convection in a stratified medium, microscopic
heat transport, viscous dissipation, and nuclear burning. Presently,
there is no obvious way to 
neglect or parameterize any of these processes in numerical
simulations without risking qualitatively wrong results. 

\section{Phase 2: The flamelet regime}
\label{flamelet}
The general features of laminar thermonuclear carbon and oxygen flames at
high to intermediate densities were described in detail by Timmes \&
Woosley \cite{timwoo92}. For our purposes, we need to know the
laminar flame speed $u_0 \approx 10^7 \dots 10^4$ cm s$^{-1}$ for
$\rho \approx  10^9 \dots 10^7$ g 
cm$^{-3}$, the flame thickness $l_{\rm th} = 10^{-4} \dots 1$ cm (defined here
as the width of the thermal pre-heating layer ahead of the much thinner
reaction front), and the density contrast between burned and unburned
material $\mu = \Delta \rho/\rho = 0.2 \dots 0.5$ (all values quoted
here assume a composition of $X_{\rm C} = X_{\rm O} =
0.5$). The thermal expansion parameter $\mu$  
reflects the partial lifting of electron degeneracy in the burning
products,  responsible for the transformation of
pre-ignition convection into a genuine Rayleigh-Taylor (RT) problem
after formation of the flame. 

In the standard picture, buoyant bubbles of ashes rising
through the fuel create turbulent velocity fluctuations $u'$ on the
scale of their diameter 
$d$, which cascade down to smaller scales. By thermal expansion and
laminar flame propagation, bubbles grow to $d \approx 10^6 \dots 10^7$
cm during the first $10^{-1}$ s of the explosion. Subject to the gravitational
acceleration $g \approx 10^9$ cm s$^{-2}$, their terminal rise velocity is 
\begin{equation}
v_{\rm r} \approx 0.4\,(\mu\,g\,d)^{1/2} \approx 10^7 \mbox{cm s}^{-1}\,\,.
\end{equation}
Hence, $u_0/u' \approx u_0/v_{\rm r} \ll 1$ and $l_{\rm th}/d
\ll 1$. These are two necessary conditions for the flamelet regime; a
third condition relating the turbulent diffusivity on 
small scales to the thermal conductivity will be discussed in
(\ref{dist}). Since the propagation of the turbulent flame brush is
dominated by the velocity of the largest turbulent eddies, the
turbulent flame speed $u_{\rm T}$ becomes independent of $u_0$, and 
\begin{equation}
\label{u'}
u_{\rm T} \sim u' \quad \mbox{if} \quad \frac{u_0}{u'} \to 0
\end{equation}
follows from simple scaling arguments. This behavior is strongly
supported by experiments \cite{shyea96}. In our context, a
short time after flame ignition the total burning rate of the
turbulent flame brush essentially decouples from the microscopic physics
of nuclear burning and heat transport, and from there on depends only on the
hydrodynamics of buoyancy-driven turbulence. Hydrodynamical
simulations of the flamelet phase in SN Ia explosions make use of this
argument (see below).  

Given the smoothness of the flame on scales $l_{\rm th}$ and the
existence of large perturbations on scales $d \gg l_{\rm th}$,
there must be an intermediate scale $l_{\rm cutoff}$ corresponding to
the transition 
between both regimes. In turbulent chemical combustion, where
turbulent velocities generally scale according to Kolmogorov scaling, $u'(l)
\sim l^{1/3}$, this 
scale is known as the ``Gibson length'' \cite{pet88}. One can find an
estimate for $l_{\rm cutoff}$ by assuming that the flame surface is
only affected by eddies that turn over at least once during their
flame crossing time: $\tau_{\rm eddy} \equiv l_{\rm cutoff}/u'(l_{\rm
cutoff}) =  l_{\rm cutoff}/u_0$, and therefore $u'(l_{\rm cutoff}) =
u_0$. For $u' \equiv u'(d)$ and Kolmogorov scaling, the cutoff scale
for flame surface perturbations is $l_{\rm cutoff} = d (u_0/u')^3$. 

The same argument holds true for flames in SN Ia explosions, except
that here the turbulence that deforms the surface is created by
buoyancy of the hot burning products and is not, as in most
terrestrial experiments, generated by a grid. As a result of the
parallel cascades of kinetic and potential energy (due to the presence
of both velocity and density perturbations), the resulting
velocity spectrum differs from Kolmogorov scaling. Rather, buoyancy-driven
turbulence conforms to Bolgiano-Obukhov scaling \cite{boror97}, 
\begin{equation}
u'(l) \sim l^{3/5}\,\,.
\end{equation}
The cutoff scale relevant for the analysis of the flamelet regime in
SN Ia explosions is thus 
\begin{equation}
l_{\rm cutoff} = d \left(\frac{u_0}{u'}\right)^{5/3}\,\,.
\end{equation}  
Below the cutoff scale, the flame surface is smooth. Buoyancy does not
couple to the turbulent kinetic energy cascade on these scales due to
the absence of density fluctuations. Hence, the velocity spectrum
turns over to Kolmogorov scaling at $l_{\rm cutoff}$,
continuing down the viscous microscale.
 
We can use this piece of information to specify the requirements for
hydrodynamical simulations. As long as  $l_{\rm cutoff}$ is
unresolved, the propagation velocity of the numerical ``flame brush'',
$u_{\rm T}(\Delta)$, is not equal to $u_0$; in fact, (\ref{u'})
suggests that it becomes independent of $u_0$ in the case of strong
turbulence.  One possible alternative is to employ
the rough approximation $u_{\rm T}(\Delta) \approx u'(\Delta)$, where the
turbulent velocity on the grid scale $\Delta$ can be extracted from a
subgrid model for the turbulent cascade \cite{niehil95}. However, with
typical values for $u' \sim 10^7$ cm s$^{-1}$ and $d \approx 10^7$ cm,
$l_{\rm cutoff}$ is not
microscopically small at high densities ($\sim 10^4$ cm for
$\rho \approx 10^8$ g cm$^{-3}$), and full resolution of this scale
is within reach of future simulations. In this case, the use of $u_0$
for the flame propagation rate is justified. The unphysical
situation associated with not resolving the viscous cutoff, of course,
remains.   

As a final remark on the validity of Gibson scaling, $l_{\rm cutoff}
\sim (u_0/u')^3$ in Kolmogorov turbulence, the turnover to smoothness
at this scale has not yet
been confirmed experimentally or by direct simulations. This fact has
been attributed to flame-turbulence interactions by thermal expansion
that are absent in the passive-surface picture employed above. In
white dwarf explosions, partial degeneracy of the burning products
results in a much lower degree of thermal expansion than in chemical
flames, where $\mu$ is generally of order unity. Therefore, the
passive-surface approximation is, compared to terrestrial flames, much
less restrictive. For purely passive flames, Gibson scaling was
clearly demonstrated in a simple discrete model for turbulent flamelet
combustion \cite{nieker2} that achieved spatial resolution of the
flame surface over four orders of magnitude.

Assuming that buoyancy effects dominate the turbulent flow that
advects the flame, the passive-surface framework obviously neglects
the additional stirring caused by thermal expansion within the flame
brush itself, accelerating the burnt material in random directions.  Both the
spectrum and cutoff scale may be affected by ``active'' turbulent
combustion \cite{niewoo97}. Although the small expansion coefficient
$\mu$ indicates that the effect is weak compared to chemical flames, a
quantitative answer is still missing. 

\section{Phase 3: The distributed burning regime and possible
deflagration-detonation-transition (DDT)}
\label{dist}
As the density of the white dwarf material declines and the laminar flamelets
become slower and thicker, a point is reached where turbulence
significantly alters the thermal flame structure. This marks the end
of the flamelet regime and the beginning of the distributed
burning, or distributed reaction zone, regime. In order to
find the critical density for 
the transition between both regimes, we need to formulate a specific
criterion for flamelet breakdown. 

In the combustion literature, the Karlovitz number $Ka = l_{\rm th}/l_{\rm
visc}$, where $l_{\rm visc}$ is the viscous cutoff scale,
is commonly used to
characterize the transition \cite{lewel61}. The flamelet regime
corresponds to $Ka <
1$, while $Ka > 1$ implies the existence of turbulent eddies smaller
than the flame thickness, which is often interpreted as the onset of
distributed burning. An alternative definition of the
Karlovitz number compares the dissipation and reaction time scales,
$Ka = \tau_{\rm b}/\tau_{\rm visc}$ \cite{brad93}. As is readily seen by
inserting $\tau_{\rm b} = \tau_{\rm th} = l_{\rm th}^2/\kappa$ and $\tau_{\rm
visc} = l_{\rm 
visc}^2/\nu$, where $\kappa$ is the thermal diffusion coefficient and
$\nu$ is the kinematic viscosity, these definitions are only
equivalent for Prandtl numbers $Pr = \nu/\kappa$ near unity. In contrast
to chemical flames where $Pr \approx 1$ is usually a reasonable
approximation, the Prandtl number in the degenerate medium of a
Chandrasekhar mass white dwarf is much smaller, $Pr \approx 10^{-5}
\dots 10^{-1}$ \cite{nanpet84}. The length and time scale criteria
diverge in this case, implying that the conditions $l_{\rm
th}/l_{\rm visc} > 1$ and $\tau_{\rm b}/\tau_{\rm visc} < 1$ can
coexist. If eddies smaller
than the flame thickness exist, but they are completely burned before
they can turn over, the Karlovitz criterion lacks an obvious interpretation.

A transition criterion that is independent of $Pr$ was proposed by
Niemeyer \& Kerstein \cite{nieker1}.  On phenomenological grounds, one
can argue that if the ratio of turbulent diffusivity, $\kappa_{\rm e}(l)
\sim lu'(l)$, and microscopic heat diffusivity $\kappa$ exceeds unity
on the scale $l_{\rm th}$ or below, the flamelet regime breaks
down. Since the turbulent diffusivity is an increasing function of
scale, it is sufficient to evaluate $\kappa_{\rm e}$ on the scale
$l_{\rm th}$. The diffusivity criterion can then seen to be equivalent
to 
\begin{equation}
\frac{\tau_{\rm b}}{\tau_{\rm eddy}(l_{\rm th})} \ge 1
\end{equation}
for flamelet breakdown. Dividing both time scales by $l_{\rm th}$, one finds
that it coincides with $u'(l_{\rm th}) \ge u_0$ and hence with
\begin{equation}
l_{\rm cutoff} \le l_{\rm th}\,\,.
\end{equation} 

Inserting the results of Timmes \& Woosley \cite{timwoo92} for $u_0$
and $l_{\rm th}$ as functions of density, and using a typical
turbulence velocity $u'(10^6\mbox{cm}) \sim 10^7$ cm s$^{-1}$, the
transition from flamelet to distributed burning was shown to occur at
a density of $\rho \approx 10^7$ g cm$^{-3}$
\cite{nieker1}. Intriguingly, one-dimensional SN Ia models that artificially 
invoke the onset of a deflagration-detonation-transition (DDT) after a
slow initial flame phase achieve best agreement with observations if
the DDT occurs very close to this density \cite{hoe96,nom97}. Lacking
a physical description of the DDT itself, this fine-tuning of the transition
density is unnatural. 

If, however,the conditions for DDT become more favorable in the
distributed burning regime than in the flamelet regime, the initiation
of a detonation could be naturally related to flamelet breakdown. It
was proposed \cite{niewoo97} that the 
local quenching of flamelets induced by turbulent mixing of fuel and
ashes can lead to the formation of macroscopic ``smoldering'' regions,
i.e. material that still burns, but on a much longer time scale than a
laminar flame under the same conditions. If only one such region
re-ignites in a nearly isothermal state, brought about by continuous
turbulent stirring, it can evoke a detonation. In order to re-ignite
before the material has expanded substantially, $\tau_{\rm b}$ in the
smoldering region must be smaller than the dynamical time scale for
expansion, $\tau_{\rm d} \sim 0.1$ s. On the other hand, turbulent
mixing must be sufficiently fast to homogenize the fluid over a length
scale given by the critical mass for detonation, $l_{\rm crit} \sim
(m_{\rm crit}/\rho)^{1/3}$:  if the the near-isothermal region is too
small at the time of ignition, the ensuing pressure wave fails to
trigger a self-sustaining detonation. The critical mass (or length) depends
sensitively on composition, density, and boundary conditions
\cite{kho97,niewoo97} and is not well known. However, under reasonable
assumptions the ordering of time scales $\tau_{\rm eddy}(l_{\rm
crit}) < \tau_{\rm b} < \tau_{\rm d}$ is fulfilled \cite{niewoo97}.

Even if the probability 
for a single DDT were very small, this may be compensated by the
(potentially) large number of critical masses in the flame
brush. Detailed knowledge of 
$m_{\rm crit}$ under various conditions, determined by direct
numerical simulations of turbulent smoldering fluid regions followed through
re-ignition, may eventually enable us
to estimate the global probability of a DDT in the distributed burning regime.

\section{Summary and conclusions}
The main goal of this essay is to classify the phases of the
thermonuclear explosion of a 
Chandrasekhar mass white dwarf in terms of the burning
regimes experienced by the turbulent combustion front. By comparing the various
length and time scales characterizing the flame, three major regimes
can be identified: the
``smoldering'' regime prior to the dynamic part of the explosion,
followed by the flamelet regime where the flame structure is
microscopically laminar but corrugated on large scales, and finally
the distributed burning regime, occurring when turbulent transport on
the scale of the flame thickness begins to dominate over microscopic
heat diffusion. 

Of these three, the flamelet regime is probably the most studied and
best understood. As long as the ``passive surface'' and ``thin flame''
assumptions hold (as specified in the main text), the turbulent flame
speed decouples from the physics 
of nuclear burning and microscopic heat transport. Instead, it scales
with the velocity of turbulent eddies, which can be estimated using
the rise velocity of hot bubbles and the Bolgiano-Obukhov spectrum of
buoyancy-driven turbulence. The full problem
involving feedback of thermal expansion on the turbulence (``active
combustion''), time-dependent gravitational acceleration, freezing-out of
the large scale eddies due to expansion of the star, and parallel
cascading of kinetic 
and potential energy continues to be accessible only to extensive
numerical simulations; however, the orders of magnitude of the
governing parameters are probably already known to reasonable accuracy.

The smoldering and ignition phase, on the other hand, is only poorly
understood. Regarding the evolution of the explosion itself, the most
important information we would like to extract is the initial shape
and location of the flame front. Likewise, the question whether the
flame is turbulent right from the beginning or makes a transition to
turbulence from an initially smooth state is still
unanswered. Detailed numerical experiments that fully resolve the
microphysical transport and burning processes are required to address
this subject.

Most important from the point of view of SN Ia modeling is the
transition from flamelet to distributed burning as the density drops
below $10^7$ g cm$^{-3}$, as this may present a physical mechanism for the
onset of a delayed detonation. Confirmation of this mechanism would
solve one of the outstanding 
fine-tuning problems of one-dimensional SN Ia models that, if the
turbulent flame speed and the density for the
deflagration-detonation-transition (DDT) are suitably chosen, agree
well with observed SN Ia spectra and lightcurves. Further
understanding of the physics of re-ignition is also needed to decide
whether the DDT occurs during the first expansion phase or after one
or several pulsations. Since the proposed
mechanism for DDT involves local flame quenching and re-ignition in a
homogenized state, the set-up for hydrodynamical simulations is very
similar to the initial ignition problem, only at a higher degree of
background turbulence and lower density. Clearly, the dynamics of a
turbulent, slightly sub-critical medium warrants closer
study. 

Given the uncertainties in the ignition and DDT processes, one can
easily find possible explanations that account for inhomogeneities
among the observed SN 
Ia events. However, lacking a detailed understanding of the underlying physics,
it is difficult to evaluate how sensibly the
global outcome of the explosion depends on minor variations of the
initial conditions or the transition criteria for burning regimes.

\section*{Acknowledgments}
I would like to thank S.E. Woosley, W. Hillebrandt, and A.R. Kerstein
for their collaboration and insights into this subject, and
J.W. Truran, R. Rosner, and D. Lamb for valuable discussions. I also wish to
thank A. Mezzacappa and his team for organizing an enjoyable
conference. This research has been supported in part by DOE contract
no. B341495.

\end{document}